# Carrier-Grade Anomaly Detection Using Time-to-Live Header Information

Quirin Scheitle, Oliver Gasser, Paul Emmerich, Georg Carle
Chair of Network Architectures and Services
Technical University of Munich (TUM)
{scheitle | gasser | emmericp | carle}@net.in.tum.de

## ABSTRACT

Time-to-Live data in the IP header offers two interesting characteristics: First, different IP stacks pick different start TTL values. Second, each traversed router should decrement the TTL value. The combination of both offers host and route fingerprinting options. We present the first work to investigate Internet-wide TTL behavior at carrier scale and evaluate its fit to detect anomalies, predominantly spoofed source IP addresses. Using purpose-built software, we capture 2 weeks of raw TTL data at a 40 Gbit/s Internet uplink. For further insight, we actively measure observed hosts and conduct large-scale hitlist-based measurements, which yields three complementary data sets for IPv4 and IPv6. A majority (69% IPv4; 81% IPv6) of passively observed multi-packet hosts exhibit one stable TTL value.

Active measurements on unstable hosts yield a stable anchor TTL value for more than 85% of responsive hosts. We develop a structure to further classify unstable hosts taking, for example, temporal stability into account. Correlation of TTL values with BGP data is clear, yet unpredictive.

The results indicate that carrier-grade TTL anomaly detection can yield significant insights in the following categories: First, the method can flag anomalies based on TTL observations (yet likely at a difficult false positive/false negative trade-off). Second, the method can establish trust that a packet originates from its acclaimed source.

## 1. INTRODUCTION

Attacks on Internet services and the underlying network infrastructure are a constant threat. Attackers often conceal their identity by using a forged source IP address. This concept is called *IP spoofing*, and is associated with malevolent intent. Spoofed IP addresses are frequently used in so-called Distributed Denial of Service (DDoS) attacks.

Using TTL characteristics for spoofing detection has been suggested by Jin et al. in 2003 [14], however their design and evaluation is based on traceroute measurements and synthetic traffic. Even more recent work, such as by Mukaddam et al. [18] in 2012 is based on small, IPv4-only, active data sets and lacks evaluation against traffic from real users.

In this work, we establish ground work for a carrier-grade traffic filter based on TTL characteristics to (a) detect IP spoofing and (b) validate packets as likely not being spoofed. We provide insight into Internet-wide TTL behavior by analyzing and correlating complementary data sets from passive and active measurements on both IPv4 and IPv6. We assess stability on host and subnet level and correlate *Hop Counts* (the likely router path length) with BGP path lengths.

Our key contributions are:

- *Carrier-grade* passive TTL capturing and analysis for IPv4 and IPv6 (Section 5)
- Dissection of TTL-unstable IP addresses into tractable groups (Section 5.3)
- Correlation of Hop Counts with BGP data (Section 6)
- Correlation of passive data with both immediate and bulk active measurements (Sections 7, 8)
- Sound data points for key design choices to build a carrier-grade TTL filter (Sections 5, 7, 8)
- Publication of data and source code (Section 3.2)

The remainder of this paper, including highlighted findings, is organized as follows: Section 2 discusses related work, and Section 3 describes our methodology, data sources, and reproducibility of our research. Section 4 details the removal of anomalies from our passive data set before evaluation. In Section 5, we evaluate the refined passive data set in depth, assessing TTL and Hop Count distributions and stability. We define a multi-layered decision tree to group IP addresses based on their various levels of TTL stability. These groupings and their corresponding sizes form an important data base for designing a TTL-based anomaly filter. In Section 6, we correlate our passively captured data set with BGP data. We find a clear correlation between AS path length and Hop Count of a subnet. However, this correlation is too noisy to be used for prediction of a "correct" Hop Count for a given subnet. In Section 7, we evaluate our *pingback* data set, gained from sending ICMP Echo Request packets to all IP addresses observed during our passive capture. These packets were sent within minutes of a passive observation and typically offer a stable TTL value for an IP address. In Section 8, we evaluate TTL characteristics for Internet-wide active scans, which offer stable TTL values per IP address and similar Hop Count characteristics (despite a different TTL start value distribution). Section 9 discussed ethical considerations undertaken by us, and Section 10 synthesizes results from the various data sets and presents promising paths for future work.

## 1.1 Background on IP Spoofing and DDoS

IP spoofing is associated with malevolent intent and hides an attacker's IP address. Yaar et al. [29] associate IP spoofing with two types of commonly observed DDoS attacks:

**SYN Flooding:** An attacker sends large quantities of TCP SYN packets to an individual destination, typically from spoofed source addresses. This forces the victim to keep state for countless half-open TCP connections, often rendering the victim incapable of accepting genuine users.

We argue that a filter based on TTL values can typically detect SYN flooding attacks: If a TTL value for the alleged source address is known, the spoofed packet would typically fail to match it. Probabilities for attacker and victim to share the same TTL value by chance are discussed in Section 5.2. If no known TTL values for an IP address exist, we argue that (a) the amount of newly learned IP-TTL-combinations will by far exceed normal thresholds, (b) the TTL characteristics will behave differently than usual and (c) predicted TTL values (e.g., from known IP addresses in the same source network) will differ from those in spoofed packets. We find one such attack in our data set and can leverage TTL values to recognize it as a flooding attack.

**Reflection Attacks:** A reflection (or amplification) attack abuses stateless protocols. These protocols are mostly based on UDP and therefore require no handshake in order to send data and trigger a response. The response can exceed the request in size (amplification). Popular protocols which are used for amplification attacks are DNS and NTP [6]. While the reflected attack traffic would easily be filterable based on its simple characteristics (known source port and few source IP addresses), its large volume typically drowns the victim's bandwidth. While TTL filtering at the victim's side does not seem promising (the TTL characteristics of arriving traffic at the victim will seem genuine), deployment at the amplifying network likely can catch this abuse based on its TTL characteristics: If a TTL value is known for the genuine source, the forged packets are unlikely to match it. If no TTL value is known, we expect to be able to learn a genuine value from active measurements. Also, if backscatter from the victim is received, it will likely carry a different TTL value than the forged request packets.

**Carrier-Scale Mitigation Strategies:** When detecting a flooding attack towards a certain target IP address, the carrier could employ standard mitigation techniques (such as dropping the first TCP SYN packet, which genuine users will likely resend) for every source IP address observed with an anomalous TTL. For reflection attacks, a staggered mitigation scheme could, depending on the level of certainty, rate-limit or filter traffic for packets with TTLs considered spoofed. Automatic or semi-automatic deployment of such techniques at carrier level could reduce the attack surface for DoS attacks.

## 2. RELATED WORK

In this section we discuss related work in the field of TTL measurements as well as spoofing and DDoS protection.

**Using TTLs for Spoofing Detection:** The idea of leveraging inbound TTL values to detect suspicious and potentially spoofed packets has been pursued by several groups. In 2004, Jin et al. [14] proposed a solution to leverage hop count filtering against spoofed DDoS traffic. Their data collection and evaluation are based on `traceroute` measurements to a limited number of targets and synthetic traffic, which limits wider applicability. In 2007, the same authors published follow-up work [25] based on the same data and evaluation concept, but providing more ideas and details. In 2007, Gill et al. proposed RFC 5082 [12], which describes a mechanism where TTL values are used to secure communications between adjacent nodes such as peering BGP routers: by setting an initial TTL of 255 and only accepting packets with a TTL of 255 on the adjacent devices, packets from outside the subnet are effectively discarded. In 2012, Mukaddam and Elhajj [18] compared various IP attributes based on active traces obtained from CAIDA's Skitter project [16]. They correlated hop count, country, Autonomous System and round-trip time of packets. They found a tendency of IP addresses within the same AS to have similar Hop Counts. They conclude that round-trip time is an effective tool to distinguish IP addresses within the same AS. However, we are not aware of spoofing or DoS problems within the same AS. We evaluated round-trip times for our measurements, but found an even less predictive correlation then for BGP data, and hence consider round-trip time as not useful for large-scale TTL analysis. In 2015, Böttger et al. [4] presented work to detect amplification attacks, suggesting comparison of passively observed TTL values with active measurements.

Our work significantly extends existing work by analyzing Internet-wide active and passive data, a prerequisite to evaluate carrier-grade feasibility.

**Generic TTL Research:** In 2006, Zander et al. [30] investigated a possible use of the TTL header for covert channels. They found flows to be very stable, and if TTLs are changed, then typically only by ±1. In 2014, Huffaker et al. [13] used a conjunction of TTL and Hop Count to assess whether two given hosts are likely geographically co-located.

**IP Spoofing & DDoS Protection:** Yaar et al. [29] propose StackPI, a traffic marking solution requiring cooperation of upstream routers. A variety of such solutions has been proposed, however they require wide-spread deployment and packet marking to work. BCP 38 (RFC 2827) [10] proposes restricting downstream networks to the allocated source addresses. Wide-spread deployment of BCP 38 would effectively hinder IP spoofing, however since its proposal in 2000, IP spoofing attacks continue to be common occurrences. Moore et al. [17] propose different flow metrics (but not TTLs) to distinguish DDoS attack traffic (including spoofed packets) from normal traffic. Wang et al. [26] propose router-level SYN flood attack detection by monitoring TCP connection establishment patterns, e.g., abundance of SYN-RST packets. In 2015, Ryba et al. [24] published a survey about amplification attack types and possible detection and mitigation strategies.

## 3. METHODOLOGY & DATA SOURCES

This section describes the approach and methodology applied to understand the nature of real-life TTL distributions. Our setup consists of three components:

**Passive Capturing:** A passive capturing machine extracts data from packets sent into the Munich Scientific Network (MWN), which connects roughly 100,000 researchers and students to the Internet through a 40 Gbit/s link. Using Intel X520-T2 cards, DPDK [1] and MoonGen [9], we efficiently extract and store the tuple of <external IP address, protocol, external Port, internal Port, TTL, timestamp> for observed incoming packets. Hardware requirements are relatively moderate, requiring 4 cores at 2.6 GHz and ≈5 GB of RAM. 2 weeks required ≈11 TB of storage.

**Active Measurements:** We conduct active measurements based on "pingback" and "hitlist" methodologies:

*pingback:* An active measurement machine receives target IP addresses from the passive capturing machine within a few minutes after an observation and conducts ICMP echo request measurements to observed IP addresses, excluding blacklisted, not routable, not assigned, and IANA special ranges. To limit third-party impact, we scan every IP address only once in every 6 hours. We use an IPv6-extended version of zmap [8] from [11]. Figure 1 depicts the pingback measurement setup.

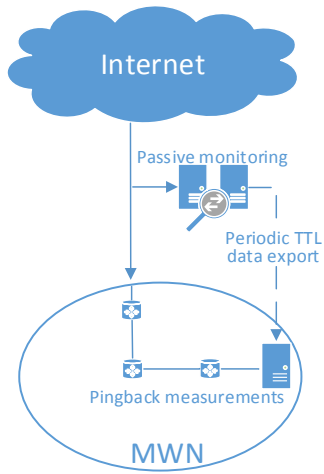

**Figure 1:** Measurement setup: Passive monitor observes packets from Internet, triggers pingback from active measurement machine.

*hitlist:* In addition to measurements on passive observations, we conduct ICMP echo request measurements on the entire IPv4 address space (also here excluding our blacklist, not routable, bogon and IANA special ranges) and towards an IPv6 hitlist according to Gasser et al. [11]. We use ICMP echo request scans according to insights of Gasser et al. [11].

**BGP Capturing:** A route server frequently stores BGP data to allow analysis for routing changes and AS path length comparison. This route server receives its updates from the same source as the router at our observation point.

### 3.1 Transformation of TTL to Hop Count

Throughout this paper, we coin TTL as the raw value observed in the IP header field according to RFC 791 (IPv4, "Time to Live") [23] and RFC 2460 (IPv6, "Hop Limit") [7]. These fields are 8-bit integers, with a possibly range from 0 to 255. For insight into the network path between source and target, it is of interest to estimate the *Hop Count*, i.e., the number of hops traversed. For estimating this, observed TTL values are mapped back to starting values such as 32, 64, 128 or 255. A frequently cited study by Paxson [21] states that most (but not all) routes in the Internet (in 1997) had less than 30 hops, but also concludes that the Internet's core has grown beyond 30 hops and hence initial TTL values larger than 30 should be used. This is confirmed by our data: 1.28% of IPv4 addresses and 0.01% of IPv6 addresses exhibit TTL values outside the range of less than 32 hops below common start values. Hence, transformation of TTLs to Hop Counts can be faulty and we therefore synonymously use the term *Estimated Hop Count*.

### 3.2 Reproducibility

We encourage other researchers to validate, reproduce and advance our research using our data. All figures and tables and certain numbers in the text in this paper are linked to a website containing respective raw data and source code to create the data representation. You can click the objects directly or browse the list on our website [2].

## 4. PREPARATION OF PASSIVE DATA SET

This section outlines details of capturing and pre-processing the passive data set. The capture period extends from February 3, 2016 through February 16, 2016. The raw dataset holds a total of 497 billion IPv4 and 37 billion IPv6 packets (≈11 TB). These packets are spread across a total of 256 million IPv4 and 2.1 million IPv6 IP addresses.

### 4.1 Aggregation into Time Bins

To allow for efficient processing of our passive data set, we create a version that aggregates data into 10 minute bins. Each bin contains the observed IP addresses and their respective TTLs. At a fraction of the raw data size, this data format allows for faster evaluation where this level of detail is adequate.

### 4.2 Traffic Statistics and Anomalies

Our first analysis aims to identify anomalies in the data set that could distort further analysis.

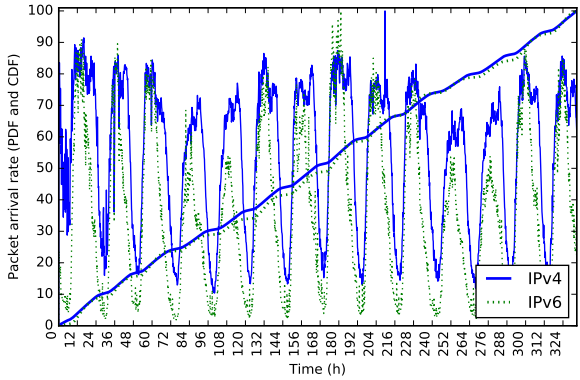

**Figure 2:** Passive Data Set: Time plot of packet arrival rate shows nocturnal pattern, ECDF disproves spikes to be of significance.

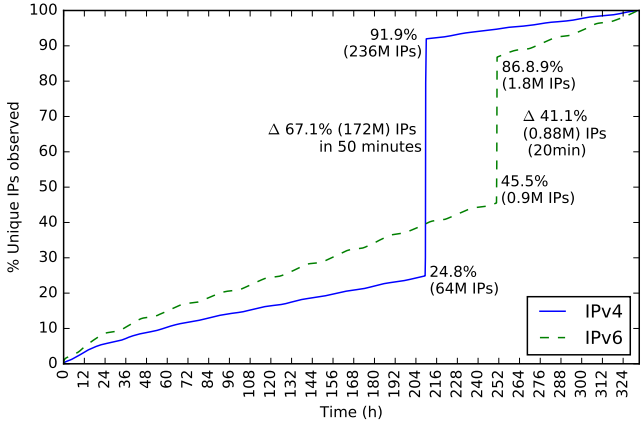

**Figure 3:** Passive Data Set: IP address accumulation shows two major step-changes of >40% each.

**Packet arrival rate** (see Figure 2) shows a nocturnal pattern, with 3AM load at ≈20% of 3PM load, and some small, but insignificant peaks. Peaks were ≈751,000 packets per second for IPv4 and ≈87,000 pps. for IPv6.

**IP address accumulation** quickly reveals major peaks for IPv4 and IPv6 (cf. Figure 3): in very short time intervals (<1h), the number of newly learned IP addresses sums up to more than 40% of all learned IP addresses. Remarkably, these peaks do not go along with significant peaks in packet arrival rate.

**IPv4 anomaly:** 160M out of the 172M IP addresses in the IPv4 anomaly are observed only during the anomaly and only with 1 packet. A further breakdown reveals that 99.9% of the 160M IP addresses sent packets towards port tcp80 within the MWN (the internal destination IP was not captured for privacy reasons). Further analysis of these packets shows unusual TTL characteristics: 99.95% of packets arrive with one of five distinct TTL values. Figure 4 shows packet arrival rates for these TTLs during the anomaly.

Geolocation of IP addresses per TTL value (using the ip2location February country-level database [3]) maps IP addresses for every distinct TTL to at least 240 different countries. As it is highly unlikely for IP addresses from 240 or more countries to arrive at our router with exactly the same TTL value, we classify this anomaly as an TCP flooding attack with spoofed source addresses. The five distinct TTLs speak for 5 different hosts carrying out a distributed attack. This underlines how flooding attacks can be recognized by their anomalous distribution of TTL values. A sophisticated attacker could try to imitate a more normal TTL distribution. With a peak of ≈464,000 packets per second and a minimum TCP packet size of 64 bytes, the attack traffic peaks at $\gtrapprox$240 Mbit/s. We exclude this anomaly from our data set by removing the affected time bins.

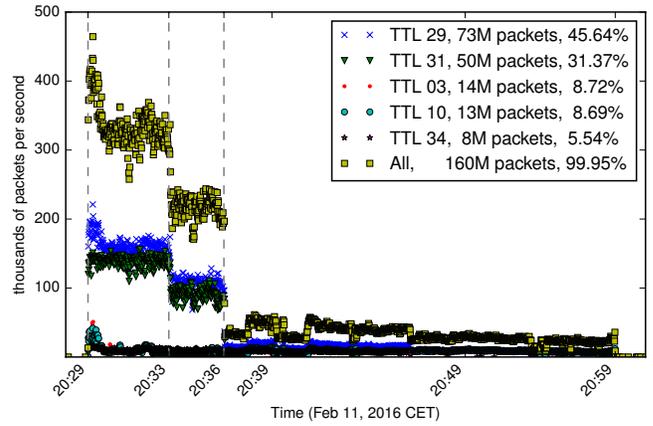

**Figure 4:** Packet rates during IPv4 anomaly.

**IPv6 anomaly:** We correlate the IPv6 anomaly with an outgoing SSL scan. This anomaly amounts to ≈41% of IPv6 addresses observed. While the return traffic from this scan is very likely genuine and carries correct TTL values, we decide to exclude this anomaly from our data set as it would include a large bias of arbitrary actively scanned hosts into our passive data set.

**Section Summary:** We analyze basic characteristics of our passive data set and find two anomalies. Using TTL characteristics, we can classify one as flooding attack. We remove both anomalies and create the refined passive dataset, which holds 85.9 million IPv4 addresses and 1.62 million IPv6 addresses. At the time of measurement, those addresses were diversely located in 240 countries (IPv4) and 143 countries (IPv6), with some centricity displayed by Germany being ranked third (IPv4) and first (IPv6).

## 5. REFINED PASSIVE DATA SET

In this section, we analyze the refined passive data set for various TTL characteristics.

### 5.1 TTL and Hop Count Distribution

First, we investigate this data set for its raw TTL distribution and its Hop Count distribution.

**TTL Distribution:** Diversity in TTL values is required to effectively distinguish between genuine and spoofed packets. If we were to observe only 2-3 TTL values, probability for an attacker to randomly share the correct TTL value of a victim would be very high. To investigate TTL diversity, we first look at its raw distribution.

Figure 5 shows the ECDF of TTLs for IPv4 and IPv6, and Table 1 shows the top 10 TTL values per for IPv4 and IPv6. We display the distributions for two different ways of counting: one based on IP addresses, where every IP can contribute every distinct TTL value once, and one packet-based, where every packet can contribute its TTL value once. Both ways of counting show the same structural behavior; however, in packet-based counting, the TTL start value of 64 gains a large share from 255. This confirms that 64 and 255 are generally common TTL start values, and that 32 and 128 are common start values for IPv4.

**HC Distribution:** We next convert the raw TTL values to Estimated Hop Counts as discussed in Section 3.1. Figure 6 shows that the majority of paths exhibit Hop Counts that indicate a path length below 16 hops (96% for IPv6, 74% for IPv4). Only 0.01% (IPv6) and 1.34% (IPv4) have Hop Counts larger than 32. This is important to keep in mind for further analysis based on Hop Counts. Figure 6 also shows

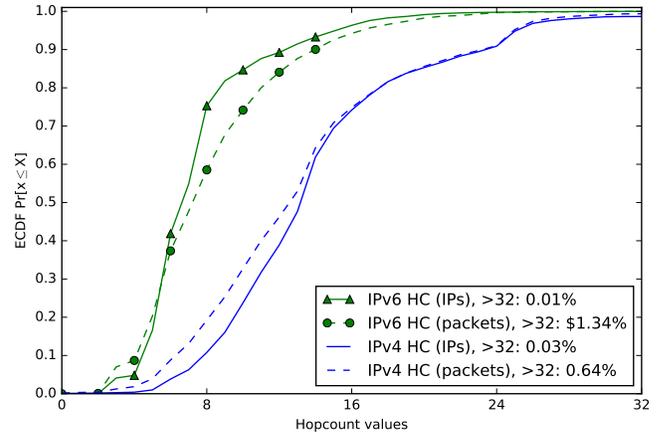

**Figure 6:** Hop Count distribution for Refined Passive Data Set. IPv6 paths are shorter than IPv4 paths, a small number of paths is longer than 32 hops.

|   | TTL | | Hop Count | |
|---|---|---|---|---|
| $n$ | IPv4 | IPv6 | IPv4 | IPv6 |
| 0 | 2.95% | 9.01% | 6.59% | 14.60% |
| 1 | 8.05% | 21.08% | 18.60% | 36.29% |
| 2 | 12.46% | 29.60% | 29.51% | 54.74% |

**Table 2:** $P_{collision}$ for TTL and Hop Count. $n$ indicates the $\pm$ value a filter would accept. Even a $\pm 1$ acceptance configuration would quickly create a large false positive acceptance range.

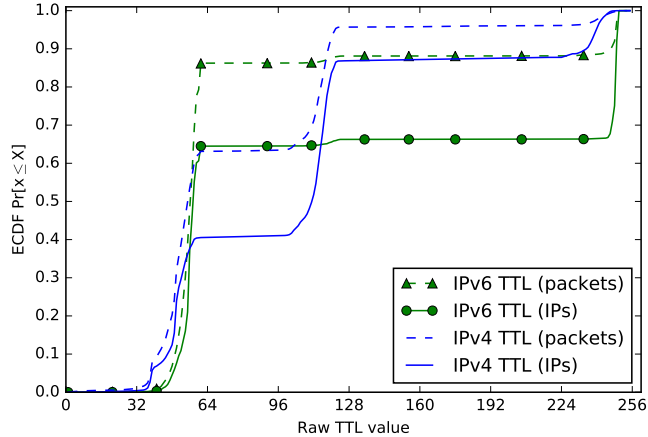

**Figure 5:** Raw TTL Distribution in Refined Passive Dataset. Bias towards start value 64 when counting per packet, bias towards start value 255 when counting per IP. Start value 128 is common for IPv4, but uncommon for IPv6. Start value 32 is still relevant for IPv4, although already considered too small by Paxson in 1997 [21].

| Rank | TTLv4 | HC[1] | %IPv4 | TTLv6 | HC[1] | %IPv6 |
|---|---|---|---|---|---|---|
| 1 | 50 | 64−14 | 8.40% | 249 | 255−6 | 17.01% |
| 2 | 117 | 128−11 | 5.05% | 56 | 64−8 | 16.56% |
| 3 | 118 | 128−10 | 4.76% | 58 | 64−6 | 9.34% |
| 4 | 114 | 128−14 | 4.64% | 57 | 64−7 | 8.19% |
| 5 | 115 | 128−13 | 4.29% | 250 | 255−5 | 7.91% |
| 6 | 116 | 128−12 | 4.28% | 59 | 64−5 | 5.01% |
| 7 | 51 | 64−13 | 3.53% | 55 | 64−9 | 4.96% |
| 8 | 39 | 64−25 | 3.13% | 248 | 255−7 | 4.18% |
| 9 | 119 | 128−9 | 3.10% | 61 | 64−3 | 3.85% |
| 10 | 49 | 64−15 | 3.09% | 247 | 255−8 | 3.15% |
| Sum Top 10: | | | 44.27% | | | 80.16% |

[1]: Displayed as estimated start value minus estimated Hop Count.

**Table 1:** Top 10 TTL values in Refined Passive Data Set, count by IP address: IPv6 offers less TTL spread and shorter Hop Counts.

that the vast majority of IPv6 paths is shorter than IPv4 paths (under the assumption that IPv6 devices do not pick, for example, a start value of 68 instead of 64). We find the Hop Count distribution to be structurally in line with the findings of Huffaker et al. [13, Figure 1] for IPv4 routers. However, comparison of raw TTL values shows a bias towards a start value of 255 in this data set, suggesting that routers frequently choose 255 as a start value.

### 5.2 Collision Probability

This section aims to assess the probability of two random IP addresses (e.g., victim and attacker) to share the same TTL or Hop Count value by chance. This would be called a *collision*. Using the empirical TTL and HC distributions, we now the estimate the collision probability $P_{collision}$. We save the empirical probability for a new packet to carry a certain TTL or HC value in an array $p[TTL/HC]$. Equation 1 then estimates the chance for two random hosts to share exactly the same or two $n$-adjacent TTL/HC values.

$$P_{collision} = \underbrace{\sum_{i=0}^{255} p[i]^2}_{\text{Same TTL/HC}} + \underbrace{\sum_{i=n}^{255} p[i] \cdot p[i-n] + \sum_{i=0}^{255-n} p[i] \cdot p[i+n]}_{\pm n \text{ TTL/HC}} \quad (1)$$

Table 2 shows $P_{collision}$ for both TTL and HC mode, based on IPv4 and IPv6 empirical distributions.

The probability of a collision climbs to 34.67% in case of IPv6 and HC$\pm 2$ operation, indicating that a too loose filter would quickly falsely permit high ratios of spoofed packets. Please also note the generally higher collision probability for IPv6 due to its centricity towards fewer raw TTL and Hop Count values.

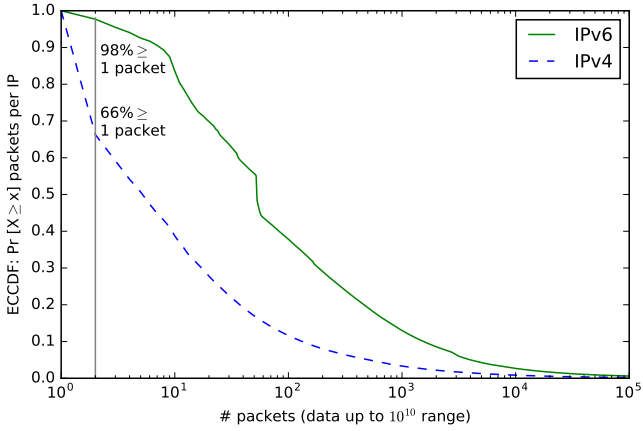

**Figure 7:** Packets per IP address in Refined Passive Dataset. IPv6 addresses have more packets, and one third of IPv4 addresses have only 1 packet.

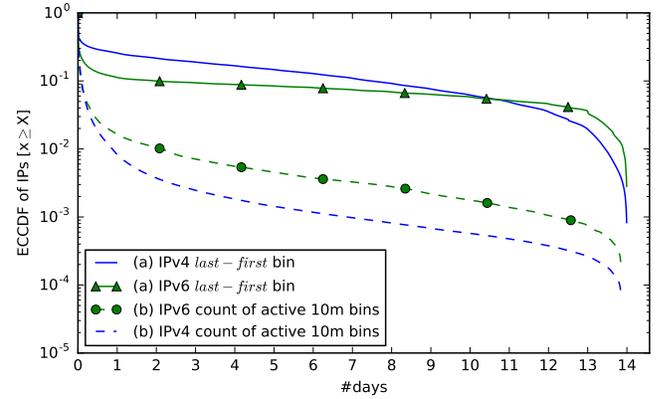

**Figure 8:** Activity Duration of IP addresses in Refined Passive Data Set: IPv4 addresses with longer duration and more active bins for IPv6. Both metrics show large portions of addresses with very short duration.

| IP duration | %IPv4 | %IPv6 |
|---|---|---|
| $\leq$ 10min | 50.1% | 59.3% |
| $\leq$ 1h | 59.0% | 73.3% |
| $\leq$ 24h | 74.4% | 88.8% |
| $>$ 24h | 25.6% | 11.2% |

**Table 3:** Selected Duration of IP address activity for Refined Passive Data Set: Tendency towards short activity, but fat tail of long-lived addresses.

Having gained understanding of time-static TTL and Hop Count distributions and collision probabilities, we now investigate longitudinal TTL and Hop Count stability per IP.

### 5.3 TTL Stability

This section aims to understand the stability of TTLs observed in our refined passive data set, i.e., whether an IP address repeatedly appears with the same or different TTL values. To properly understand TTL stability per IP address, we first need to gain understanding about two potentially influencing factors: the number of packets per IP address and the observation duration per IP address.

**Number of packets:** The number of packets can influence the validity of TTL stability for IP addresses in two ways. For one, addresses with very few packets could exhibit a more stable TTL behavior than addresses observed with many packets. In the extreme case, IP addresses observed with only one packet will obviously always be TTL-stable. Figure 7 shows the ECCDF of number of packets observed per IP: 2% of IPv6 addresses and 33% of IPv4 addresses are observed with only one incoming packet and hence do not allow for TTL stability assessment. IPv6 addresses were typically active with more packets than IPv4 addresses. While most IP addresses originated 1000 or less packets, some addresses have sent 10 billion packets or more.

However, the number of packets alone can not distinguish two very different cases: An equal number of packets could either be allocated to a single burst or be spread out equally over weeks. The latter case is more prone to TTL instability from topology changes. We hence investigate the duration of IP address observations before judging TTL stability.

**Duration:** We measure duration with two metrics: The first, observation duration, calculates the time difference between the last and the first observation of an IP address. The second counts the number of 10-minute bins that an IP address has been observed in: this metric is easy to calculate from our binned format described in Section 4.2, and insightful when used to rate sustained activity of an IP.

Figure 8 shows both these metrics as an ECCDF. We see that both duration and count of active bins show a tendency towards short observations, with more than 50% of IP addresses occurring in only one 10-minute interval. As we cut duration at observation start and end, our metrics potentially underestimate duration for some flows extending through these data points. Also, some of the IP addresses being observed for 13 or 14 days likely will be observable as virtually always active, such as root DNS servers. Table 3 quantifies the share of IP addresses for selected durations.

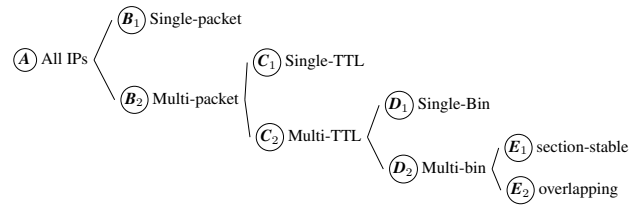

**Figure 9:** Structure to evaluate TTL stability aims towards fanning out unremarkable characteristics (such as single-packet IP addresses) early.

**Stability Evaluation:** Having analyzed the distribution of packet counts and observation durations, we now take a closer look at TTL stability. To cope with the high dimensionality of potential metrics and the high number of entries, we devise a structured decision tree, aiming to quickly categorize unremarkable behavior at an early stage. This early categorization allows for more in-depth analysis of conspicuous addresses at a later stage. Figure 9 depicts the structure of this staggered approach, Table 4 quantifies the size of the respective groups.

| Category | IPv4# | IPv4% | IPv6# | IPv6% | cf. |
|---|---|---|---|---|---|
| $A$ All | 85,986,816 | | 1,625,795 | | |
| $B_1$ Single-packet IPs | 28,986,129 | 33.71% | 37,354 | 2.30% | F7 |
| $B_2$ IPs >1 packet | 57,000,686 | 100.00% | 1,588,441 | 100.00% | F7 |
| $C_1$ IPs with 1TTL | 39,431,567 | 69.18% | 1,280,999 | 80.65% | F10 |
| $C_2$ IPs >1TTL | 17,569,119 | 30.82% | 307,442 | 19.35% | F10 |
| $D_1$ IPs with 1 active bin | 1,467,531 | 2.57% | 118,933 | 7.49% | F8 |
| IPs with 1 HC | 48,950 | 0.09% | 35,996 | 2.27% | |
| IPs with >1 HC | 1,418,581 | 2.49% | 82,937 | 5.22% | |
| $D_2$ IPs with >1 active bin | 16,101,588 | 28.25% | 188,509 | 11.87% | F8 |
| $E_1$ section-stable TTLs | 12,923,493 | 22.67% | 89,148 | 5.61% | |
| IPs with 1 HC | 1,791,556 | 3.14% | 33,517 | 2.11% | |
| IPs with >1 HC | 11,131,937 | 19.53% | 55,631 | 3.50% | |
| $E_2$ Overlapping TTLs | 3,178,095 | 5.58% | 99,361 | 6.26% | |
| IPs with 1 HC | 83,459 | 0.15% | 27,137 | 1.71% | |
| IPs with >1 HC | 3,094,636 | 5.43% | 72,224 | 4.55% | |

**Table 4:** Refined Passive Data Set. Quantitative data for TTL characteristics based on decision tree: Behavior for IPv4 and IPv6 sometimes strongly differs. Figures for certain data referenced in "cf." column.

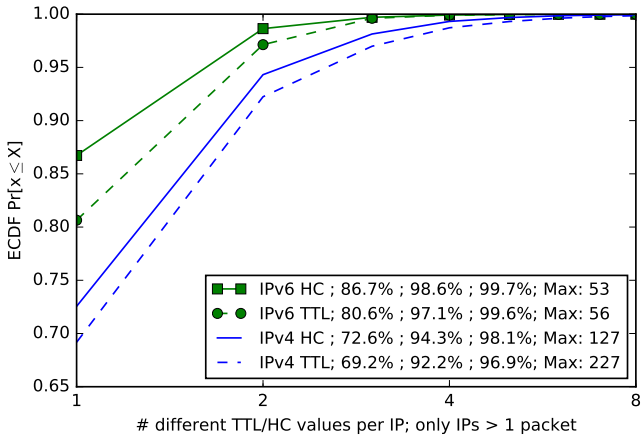

**Figure 10:** Number of different TTLs and HCs per IP for Refined Passive Data Set: Less Hop Counts per IP, IPv6 with generally less spread. Legend gives % for 1 and 2 different TTLs/HC as well as maximum value observed.

$B_1$ **- Single-packet IP addresses:** These do not allow for resilient study of TTL characteristics and are hence filtered out. It is remarkable that for IPv4, a very high number of IP addresses (33.71%, ≈29M) occurred with only 1 packet, compared to only 2.3% (≈37K) of IPv6 addresses.

$C$ **- Simple TTL stability:** We next look at stability of TTL and Hop Counts per IP. At this stage, we can already greenlight IP addresses observed with only 1 TTL ($C_1$). This includes 80.6% (IPv6) respective 69.2% (IPv4) of IP addresses. Figure 10 shows the ECDF of TTLs and Hop Counts per IP. For IPv4 and IPv6, we find fewer different Hop Count values than TTL values. IPv6 shows higher stability for both TTL and Hop Count metrics.

$D$ **- Count of Bins:** For IP addresses with several observed TTLs, we next dissect "overlapping" (several active TTLs at the same time) and "section-stable" (only one active TTL at one point in time) IP addresses. We first filter out all IP addresses that were active in only one 10-minute bin (cf. Figure 8), as these do not allow for longitudinal analysis.

$E$ **- Section Stability:** IP addresses active in several bins ($D_2$) are considered section-stable if no consecutive *mixbins* (bins with more than 1 TTL per IP) exist. This allows 1 *mixbin* for transitioning between TTLs. Hypothesis for the set of section-stable IP addresses $E_1$ is that these are, for example, dynamically assigned IP addresses at an ISP or legitimate TTL changes due to routing changes. IP addresses considered non section-stable (i.e., overlapping) are grouped in $E_2$. Interestingly, section-stable IP addresses clearly dominate (≈78%) multi-TTL, multi-bin IPv4 addresses, while having only a 50% share at IPv6 addresses. We hence argue that dynamic reassignment of IP addresses to different end hosts is much more common in IPv4 than in IPv6, a likely behavior given IPv4 address scarcity.

To conclude, a simple approach can easily legitimate the TTL behavior of 69% (IPv4) and 81% (IPv6) of addresses. While it is difficult to classify the remainder as "normal" or "not normal", we succeeded in structuring it into several clearly distinguishable groups that allow further investigation based on underlying root causes for their behavior.

### 5.4 Hop Count Stability

To further understand the cases of multi-TTL IP addresses, we conduct analysis on estimated *Hop Count* values. Certain network characteristics, such as dynamic IP assignments or DNS-resolving NAT routers, might be verified or falsified by Hop Count characteristics. We hence investigate the *Hop Count amplitude (HCA)*, defined as the maximum Hop Count minus the minimum Hop Count per IP address, as suggested by Zander et al. in 2006 [30].

Figure 11 displays ECDFs for various groups along our decision tree and offers several clear conclusions: **(a)** IPv6 *HCA* more quickly saturates than IPv4, **(b)** IP addresses with "overlapping" TTL behavior show the worst HCA and **(c)** filtering on a HC amplitude of 1 would yield 92% (IPv4) and 98% (IPv6) legitimate rates. Concluding, HCA filtering can increase the level of insight over raw TTL matching, but also increases $P_{collision}$ (cf. Section 5.2).

### 5.5 TTL Amplitude

The insights into Hop Count amplitude from the previous section raise the question on TTL amplitude across different groups of IP addresses. Figure 12 displays the ECDF for IPv4 and IPv6 TTL amplitude across the same groups of IP addresses as the Hop Count amplitude analyses in the previous section. It yields very powerful insights: TTL changes group either at several small values, likely caused by routing changes or NAT anomalies, or at one distinct very large value, likely a change of start values. We argue that combining Hop Count amplitude and TTL amplitude values in case of a TTL change can bring important insights: A high TTL

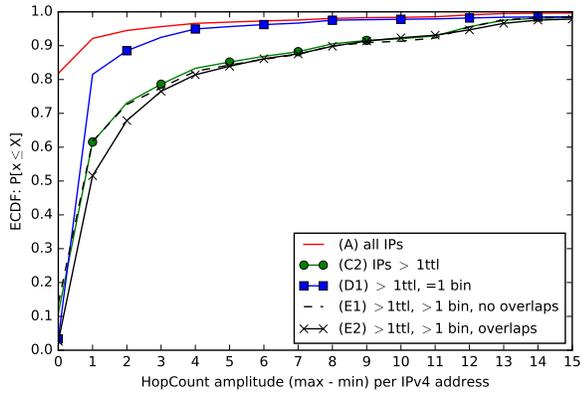

(a) IPv4 HC with more spread than IPv6 HC.

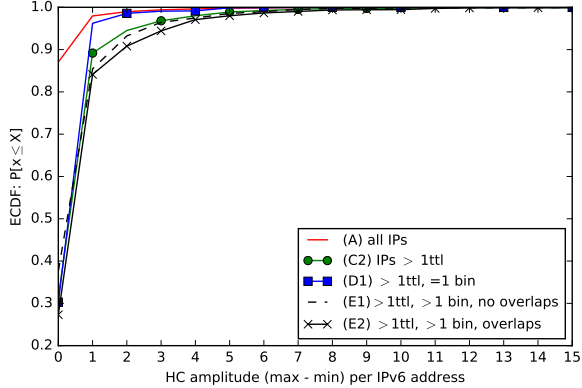

(b) IPv6 HC with less spread IPv4 HC.

**Figure 11:** Hop Count amplitudes (Max - Min) per IP for Refined Passive Data Set: Single-bin IP addresses with lower Hop Count amplitude, likely due to less probability for major routing changes in 10minute-bin. IP addresses with overlapping TTL behavior with biggest HC spread.

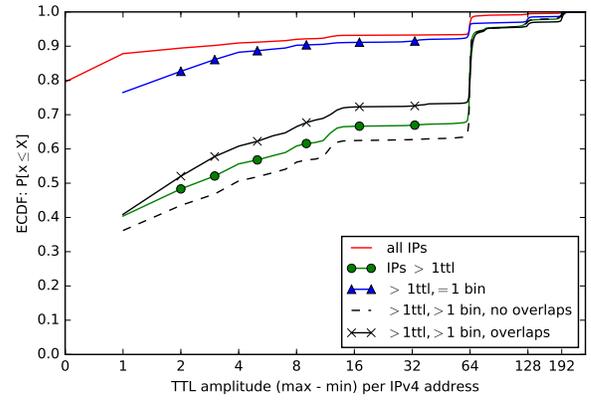

(a) IPv4: Step changes from 0 to 10 (likely routing changes) and at 64 (likely change between main IPv4 start values of 64 and 128, cf. Figure 5). ≈90% of single-bin IP addresses with small changes.

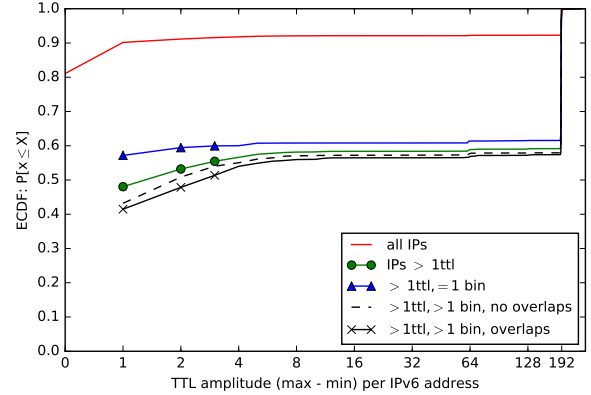

(b) IPv6: Step changes from 0 to 10 (likely routing change) and around 192 (likely change between main IPv6 start values of 255 and 64, cf. Figure 5). Single-bin IP addresses not different at IPv6.

**Figure 12:** Refined Passive Data Set: ECDF of TTL amplitude (max-min) for IPv4 and IPv6 highlights two step changes: Very small values for, e.g., routing changes or NAT gateways and very large values, likely for start value changes.

amplitude indicates a change in the start value (and hence likely the connected device), while a low TTL amplitude indicates a sole routing change. Both insights could potentially improve a TTL filter by enriching its knowledge.

### 5.6 Flow-Level Analysis

Further insight could be gathered by analyzing flow-level data. We are aware that this is a challenging undertaking at carrier-grade traffic levels. However, it might (a) provide exemplary insight in this study and (b) still be conducted at carrier-grade, if only applied to a minority of flows. For privacy concerns, our data set does not contain IP addresses in the carrier's customer cone (cf. Section 9). Hence, we are not working on flows as commonly defined by a 5-tuple, but on *4-flows*. Using the birthday problem, we estimate a lower bound for collision probability. Using an upper bound of 65,000 randomly chosen ephemeral ports per IP (an average could be estimated from [15, Table 1]), equation 2 gives the collision probability for an external IP and $n$ concurrent internal users:

$$P_{\text{flowcollision}} = 1 - \frac{n!\binom{65000}{n}}{65000^n} \qquad (2)$$

We calculate a collision risk of 7.3% for n=100 users accessing an external IP at the same time, and 99.95% for n=1000 users. As the observed network peaks at 100K users, we can assume a large share of collisions in 4-flows.

In our experiment, we found >90% of multi-TTL IP addresses to exhibit at least one 4-flow with several TTL values. Closer analysis revealed diverse characteristics, e.g., (a) outbound DNS, (b) outbound HTTPS to large sites or (c) inbound mail traffic. As this high share of routing changes or multi-TTL load-balancing seems highly unlikely, we discard further analysis of our 4-flows at this stage. We conclude that significant insight into the behavior of unstable IP addresses at carrier-grade will have to be gained from 5-flows. Future work should investigate anonymized, privacy-preserving capture of 5-tuples to allow collision-free flow tracking.

**Section Summary:** Concluding analysis of our refined passive data set, we found a majority of IP addresses to be TTL-stable and even more to be Hop Count-stable. We defined and analyzed several subgroups of unstable hosts, likely linked to network phenomenons causing these characteristics. This data stock should be of great help in designing a TTL-based anomaly filter. We next correlate our refined passive data set with BGP data.

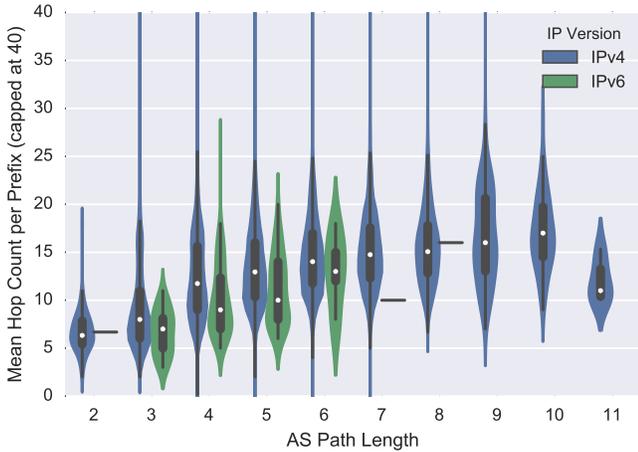

**Figure 13:** Distribution of Mean of Hop Counts per announced prefix (Y-Axis) over AS path length of announced prefix (X-Axis). IPv4 displayed on left side, IPv6 on right. Path lengths with too few data points displayed as horizontal lines (IPv6). White dot shows median, consistently lower for IPv6. Created using Seaborn library [27].

| Coefficient | IPv4 | IPv6 |
|---|---|---|
| Slope | 1.14 | 1.81 |
| Intercept | 7.74 | 2.18 |
| $R^2$ | 0.05 | 0.15 |
| P value | $< 10^{-8}$ | $< 10^{-8}$ |

**Table 5:** Linear regression parameters for Mean of Hop Counts per announced prefix compared to AS path length of Prefix. Very low $P$ value indicates strong correlation, low $R^2$ indicates low predictability.

## 6. BGP CORRELATION

We collected BGP data during the capture period, at a router receiving the same updates as the router deciding on routing on the link we observe. Using this data, we investigate whether there is a correlation between Hop Count and AS path length. We reduce the problem to a time-static problem by (a) taking a full RIB snapshot at the beginning of our capture period and (b) averaging all TTL values per announced subnet in our refined passive data set.

Figure 13 shows the distribution of mean Hop Count per announced prefix over the AS path length of the same prefix. The violin plots combine a box plot (center line) and a kernel density estimate ("violins").

Visual inspection offers several insights: (a) Hop Count seems to increase with increasing AS path length, (b) Hop Count distribution has a high standard deviation, (c) IPv6 paths are on average shorter in Hop Count length and (d) IPv6 AS path lengths show less diversity, with almost no paths outside the 3 to 6 range. We calculate linear regression coefficients for both IPv4 and IPv6 and find both to clearly show the existence of a correlation (both $p$-values $< 10^{-8}$). Albeit, this correlation suffers from low prediction power for IPv4 ($R^2 = 0.05$) and IPv6 ($R^2 = 0.18$). This is equivalent to the large spread visually observed. We conclude that AS path length and mean Hop Count per prefix correlate, but spread is too large to use this data for prediction of likely Hop Count values from AS path length. We note that these numbers potentially contain bias where routes are asymmetric. Although we record the BGP routes for outgoing packets, inbound packets might have followed a different route. For future work, analysis of the time-dynamic case might yield interesting insight into the correlations of routing changes and Hop Count changes.

**Section Summary:** We find AS path length clearly correlates with Hop Count, but is an unfit predictor due to the large spread of Hop Count values.

## 7. EVALUATION OF PINGBACK SCANS

To better evaluate the TTL values which we capture using passive monitoring, we conducted regular active measurements. These pingback measurements are performed for each IP address that we observed in the passive monitoring setup. We probe each seen IP address with an `ICMP echo request` sent by zmap less than five minutes after the IP is observed. This low delay ensures that any influence of changing Internet routes is kept minimal. To reduce the load on very active IP addresses, we probe each IP address only once every six hours. As we run measurements every minute, we use the short default zmap timeout of 8 seconds. This might miss some IP addresses with very long delays (cf. Padmanabhan et al. [19]). Our pingback measurements still yielded a rather high response rate of 46% (IPv4) and 42% (IPv6) of queried IP addresses. Table 6 displays the basic statistics about our pingback measurements.

| | IPv4 | IPv6 |
|---|---|---|
| IPs scanned | 85.9M | 1.6M |
| IPs replied | 46.12% | 41.17% |

**Table 6:** Pingback Measurements offer a rather high response rate of 46% (IPv4) and 42% (IPv6) of queried IP addresses.

The following subsections investigate TTL and Hop Count distributions, assess TTL stability within one measurement run and over the complete period, and compare TTL stability of passive and pingback data.

### 7.1 IP-Level TTL/HC Distribution

We first compare raw TTL and Hop Count distributions between the passive data set and the pingback data set. As our pingback machine is located 3 hops behind our passive observations, we add an offset of 3 to all raw pingback TTLs (no overflows occurred). Figure 14 shows the distribution of TTLs: the pingback measurements exhibit a very high share (>90%) of IP addresses with a TTL start value of 64, which is different from the distribution of our passive data set. Three hypotheses could explain this behavior: (1) a structurally different subset of IP addresses replied to our pingback measurements, (2) middle-boxes with a start value of 64 replied to our ping requests or (3) IP stacks predominantly use 64 as a start value for ICMP, but not for TCP/UDP. Figure 15 compares the Hop Count distribution from our pingback measurement to that from our passive data. IPv4 active paths seem a bit shorter, which could be caused by middle boxes replying to `ICMP echo request` packets.

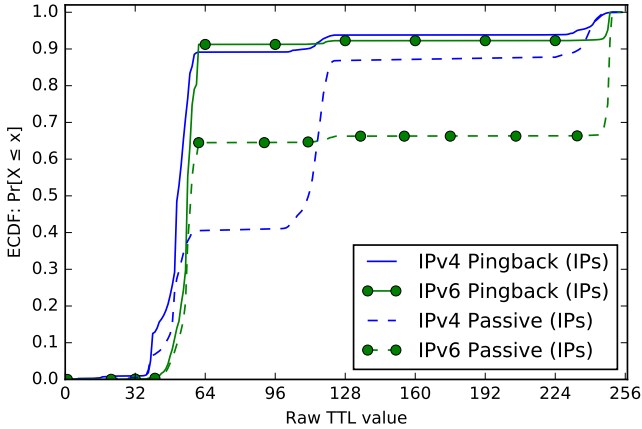

**Figure 14:** Pingback Raw TTL Distribution: ≈90% of IP addresses in Pingback measurements with a TTL start value of 64, compared to 65% (IPv4) and 40% (IPv6) in passive observations.

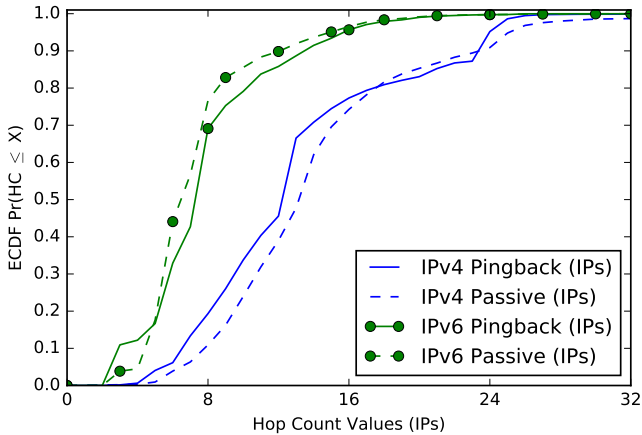

**Figure 15:** Pingback Hop Count Distribution: Pingback measurements with slightly shorter paths for IPv4, potentially caused by middle-boxes replying to `ICMP echo request` packets.

## 7.2 Stability within Measurement Run

To further analyze this phenomenon we evaluate TTL stability within one measurement run (we ran measurements every minute). This way we can identify IP addresses which behave strangely by responding with multiple TTLs even within a single minute, which are likely caused by load balancers or other middle boxes and less likely by routing changes. Tables 7 and 8 show the number of distinct TTL values of one IP within one measurement run. The vast majority (>99%) have a single TTL within one measurement run. We find 0.84% of IPv4 addresses with more than one TTL value. This is in contrast to IPv6 where only 0.09% exhibit multiple different TTLs. Generally, we can see that IPv6 shows a much cleaner picture with regard to TTL pingback stability. Some IPv4 addresses send more responses than the three request packets that we sent. One specific IP even repeatedly sent us more than 100 responses for our three request packets. This IP belongs to a Ukrainian ISP which seems to operate a NAT pool. We assume that in this case the `ICMP echo request` packet was distributed to all users behind the NAT pool. We looked at rDNS names of the 20 IP addresses with >3 TTLs and found 13 dynamic (IP encoded), 1 name server, 1 measurement machine, 2 "gateways" and 3 IP addresses without rDNS names.

| Distinct TTLs | Occurrence | Percentage |
|---|---|---|
| 1 | 87,945,276 | 99.16% |
| 2 | 729,866 | 0.82% |
| 3 | 17,319 | 0.02% |
| 4 | 10 | 0.00% |
| 5 | 3 | 0.00% |
| 6 | 4 | 0.00% |
| 7 | 1 | 0.00% |
| 8 | 1 | 0.00% |
| 42 | 1 | 0.00% |
| 103 | 1 | 0.00% |
| 106 | 1 | 0.00% |
| 107 | 2 | 0.00% |

**Table 7:** Pingback: IPv4 addresses generally respond with 1 TTL within one measurement run with some interesting exceptions. IP addresses scanned multiple times were counted multiple times.

| Distinct TTLs | Occurrence | Percentage |
|---|---|---|
| 1 | 2,493,673 | 99.91% |
| 2 | 2,329 | 0.09% |
| 3 | 9 | 0.00% |

**Table 8:** Pingback: IPv6 addresses generally respond with 1 TTL within one measurement run.

## 7.3 Longitudinal IP Stability

We further evaluate the longitudinal TTL stability of IP addresses. Therefore we define three mutually exclusive stability classes: (1) IP addresses which are stable over the complete measurement period, (2) IP addresses which are stable within measurement runs but exhibit TTL changes over the 2-week experiment, and (3) IP addresses which are already unstable within one measurement run.

Table 9 shows the distributions of IPv4 and IPv6 addresses in these three classes. We can see that over 96% of IP ad-dresses are stable over the complete measurement period. There is, however, a non-negligible portion of IP addresses which has stable TTLs within measurement runs, but exhibits differences in TTL values over the complete measurement period. This behavior could stem from routing changes or load balancers choosing different paths for new measurements. These unstable but bin-stable IP addresses are more prevalent in IPv4 than IPv6. Finally, bin-unstable IP addresses are ten times more frequent in IPv4 than IPv6. This hints at a more frequent deployment of TTL-unstable load balancers and more complex routing topologies in IPv4.

| Stability | | IPv4 | IPv6 |
|---|---|---|---|
| (1) Stable | | 96.13% | 97.93% |
| (2) Unstable | Bin-stable | 2.92% | 1.98% |
| (3) | Bin-unstable | 0.95% | 0.09% |

Table 9: Pingback: Longitudinal stability of IP addresses.

To conclude, most IP addresses are stable over the complete measurement run, IPv4 has a greater percentage of unstable TTLs than IPv6, especially when measuring within a short time span.

### 7.4 Stability compared to Passive Data

One critical question about our pingback measurements is whether these can help to establish a known, stable TTL (or Hop Count) value for an IP address that exhibits ambiguous TTL (or Hop Count) values. To assess these questions, we took the set of addresses observed with more than 1 TTL (and hence, more than one packet) in our passive observations. For those IP addresses, we evaluated the pingback results and classified them into stable, unstable or unresponsive. Table 10 shows this classification:

First, a large number of addresses simply does not reply to `ICMP echo request` packets. These values are in line with expectations and might be improved by sending in-protocol packets. Second, of the responsive packets, a majority exhibits one single TTL value. This behavior is very encouraging for the goal of building a TTL-based filter: Pingbacks to unstable hosts are possible, and typically provide a stable TTL value. This behavior can be used to infer validity of observed TTL values, for example, by using this value as a lower bound for the expected Hop Count (under the assumption that `ICMP echo request` packets would be responded to directly by a load balancer).

| | IPv4 | IPv6 |
|---|---|---|
| Multi-TTL IPs | 17.6M | 0.3M |
| Pingback: =1 TTL | 42.59% | 26.33% |
| Pingback: >1 TTL, bin-stable | 6.00% | 4.10% |
| Pingback: >1 TTL, bin-unstable | 1.08% | 0.14% |
| Pingback: no response | 50.34% | 69.42% |

Table 10: Addresses responsive to pingback measurements typically provide one single, stable TTL value. Only ≤1.1% of IP addresses bin-unstable.

**Section Summary:** In this section, we investigated TTL and estimated Hop Count distribution for the pingback measurements conducted. We found a strong bias towards a start value of 64 in the start values, but a similar Hop Count distribution. While over 50% of passively observed IP addresses would not respond to pingback measurements, those who did respond would typically provide a singular TTL value. Some IP addresses replied with more packets than were sent.

## 8. EVALUATION OF HITLIST SCANS

We scan the routed IPv4 address space, excluding our blacklist, and an IPv6 hitlist according to Section 3. Table 11 breaks down the number of packets sent and the various reply types received. Our scanning host connects to the Internet via the same uplinks which we monitored for the passive data sets. However, it is 3 hops behind the monitored link. We correct this offset in the active data set. We only evaluate `echo reply` packets to not distort our data set with error responses from routers. For those, we first evaluate raw TTL distribution to gain insight into the TTL behavior of our active `ICMP` probes compared to the passively observed traffic (which is UDP/TCP).

| Category | IPv4 | IPv6 |
|---|---|---|
| Target IPs | 2.744B | 1.962M |
| Responsive IPs | 14.20% | 64.48% |
| Multi-TTL IPs[1] | 1.07% | 0.61% |
| Packets sent | 8.232B | 5.887M |
| Packets received | 17.37% | 67.46% |
| Echo reply[2] | 87.91% | 91.85% |
| Unreachable[2] | 5.34% | 4.43% |
| Redirect[2] | 3.72% | N/A |
| Time exceeded[2] | 3.01% | 1.63% |
| Other[2,3] | 0.02% | 2.09% |

1: % of responsive IPs    2: % of received packets
3: Including, e.g., Prohibited or Reject Route.

Table 11: Hitlist Scans: Much higher response rate for IPv6 (as expected), responsive IP addresses very rarely observed with > 1 TTL.

### 8.1 IP-Level TTL/HC Distribution

Figure 16 displays the raw TTL distribution of both scanned IP addresses and passively observed IP addresses. Similar to our pingback measurements (cf. Figure 14), a TTL starting value of 64 is more common in our hitlist scans. This effect is quite interesting and could be explained by receiving replies from different devices other than the actual end hosts (such as a NAT gateway). However, converting the raw TTL values to estimated Hop Counts (cf. Figure 17) shows no significant difference between the active and passive data sets. This shows that while the raw TTL distribution for active and passive measurements differ, the difference is solely limited to different start values and does not affect the estimated Hop Count. This also means that our passive data set has no significant bias regarding the distance of the observed hosts, which speaks for a representative passive sample.

| | IPv4 | | IPv6 | |
|---|---|---|---|---|
| Rank | Amplitude | Frequency | Amplitude | Frequency |
| 1 | 1 | 90.83% | 1 | 79.28% |
| 2 | 2 | 3.40% | 5 | 9.80% |
| 3 | 3 | 1.83% | 2 | 8.71% |
| Multi-TTL IPs: | | 4,157,730 | | 7,769 |

**Table 12:** Top 3 TTL amplitudes per IP for Multi-TTL IP addresses in Active Scan indicate a typical low spread of TTL values for active scans.

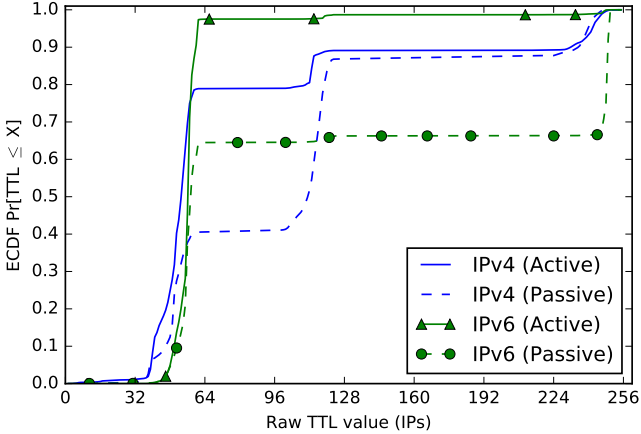

**Figure 16:** Raw TTL distribution of IP addresses: Active scans (solid lines) with larger share of a 64 start value than passive observations (dashed lines).

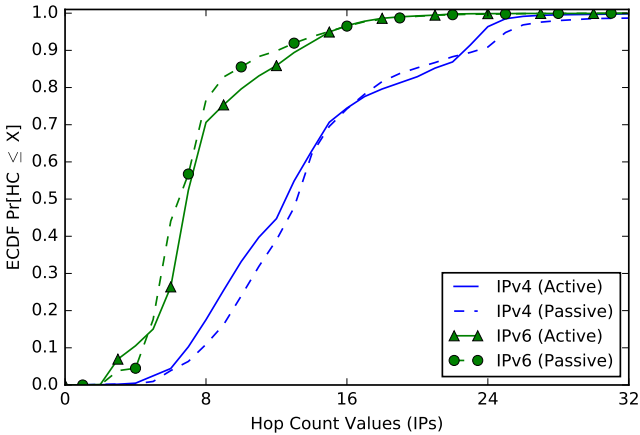

**Figure 17:** Hop Count distribution of IP addresses: IPv6 paths (dot-marked lines) shorter, no significant difference between active and passive measurements.

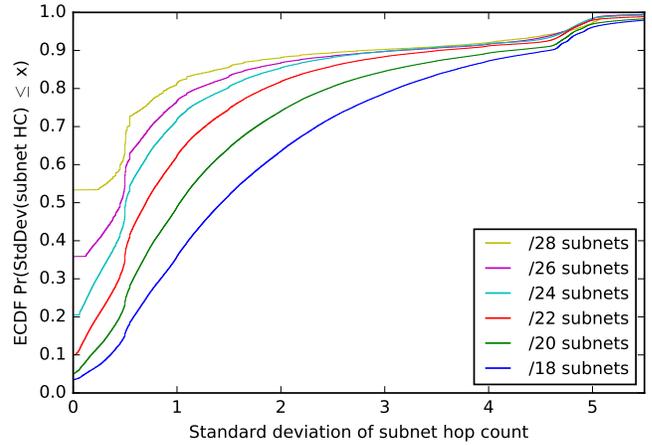

**Figure 18:** Standard deviation of Hop Counts within IPv4 subnets: Smaller subnets exhibit smaller standard deviations, 20% of /24 subnets with $\sigma = 0$.

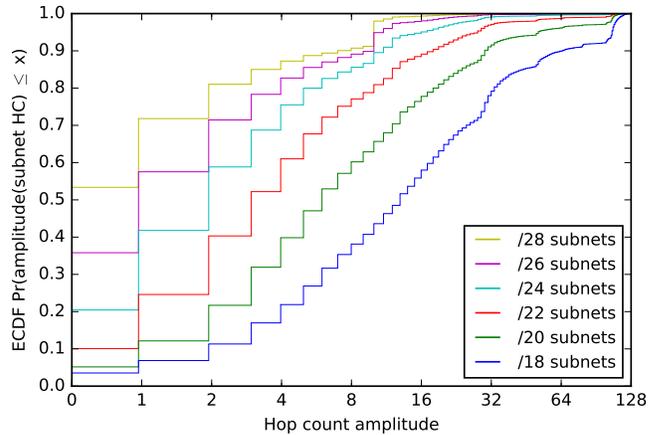

**Figure 19:** Amplitude (max-min) of Hop Counts within IPv4 subnets features same structure as standard deviation (cf. Figure 18). Grouping into a filter allowing a median $\pm 1$ would, e.g., work for 60% of /24 subnets.

## 8.2 Subnet Level TTL/HC Distribution

We next analyze TTL and estimated Hop Count distribution for various subnet sizes. Working on subnet aggregates instead of individual IP addresses could reduce memory usage in a filter and also give a-priori information about newly observed IP addresses stemming from a known subnet. We group the IP addresses observed in our active scans into equisized subnets to evaluate the feasibility of such an aggregation. This approach only yields useful results for our full IPv4 scan as the number of IPv6 hosts is very small compared to even a /64 subnet size. Future work could leverage a radix tree, suggested by Cho et al. [5] and recently applied by Plonka and Berger [22] to IPv6, to dynamically group IPv6 addresses into subnets of adaptive size.

Figure 18 shows the stability of active subnets (with at least one active host) by plotting the standard deviation of Hop Counts within subnets. A standard deviation of 0 means that a subnet is completely stable – all reachable hosts in a subnet had the same Hop Count. Note that this figure includes trivially stable subnets with only a single active host or subnets with few hosts. It is still useful for the proposed filter as aggregating subnets with few hosts is still valid and useful. The data shows that even relatively large groupings such as /24 subnets are completely stable in 20% of the cases, and /28 subnets are even stable in $> 50\%$ of cases. Figure 19 plots the difference between the maximum and minimum estimated Hop Count observed within a subnet: For a /24 subnet, for example, $\approx$60% of subnets could be grouped within a range of 3 Hop Counts (a median $\pm 1$). As expected, amplitude increases with subnet size, posing an optimization problem to find a suitable aggregation size.

The Hop Count spread can be caused by various reasons, e.g., cascades of NAT devices or splitting up a /24 subnet into layer of smaller subnets with hierarchical routing. Investigating groups of addresses with equal Hop Counts within a subnet could be an interesting course for future work. Further, data from BGP can be incorporated to determine subnet sizes. However, BGP data only represents an upper bound for the subnet size as the announced, often aggregated prefix can be split into smaller subnets within the AS. Subnet aggregation with permissive ranges will also increase the chance for a random attacker to have a correct Hop Count. For example, the previously discussed filter that would group 60% of /24 subnets to a median $\pm 1$ would suffer from a 18.6% chance for an attacker to randomly offer a correct Hop Count (cf. Table 2). Having discussed the distribution of TTL and Hop Count from our active hitlist scans, we conclude that dynamically tracking and, where needed, splitting, subnet sizes could be a feasible course of action for a filter to reduce memory need and predict likely Hop Counts for new IP addresses. However, adjusting for low false positive and negative rates could be a challenging task.

**Section Summary:** In our hitlit scans, we find responsive IP addresses to typically show stable TTL behavior. Unstable IP addresses typically have very low TTL amplitude. Raw distribution is (as for pingback) biased towards a start value of 64, while Hop Counts are similar. We aggregated for different subnet sizes and show that, depending on subnet size, significant shares of subnets show TTL-stable behavior. This confirms potential use of a dynamic subnet aggregation/splitting.

## 9. ETHICAL CONSIDERATIONS

This section outlines ethical considerations undertaken regarding active and passive measurements and data release. Our research group follows an internal multi-party approval process before any measurement activities are carried out. This approval process incorporates the proposal of Partridge and Allman [20] to assess whether the collection of data can induce harm on individuals and whether the collected data reveals private information. We draw the following conclusions from this process:

**Active Measurements** can be interpreted as an attack, resulting in investigative effort for system administrators. To minimize the intrusiveness of our active network measurements we implement the following procedures:
First, we set up a website on the scanning machines which explains our measurement activity in detail, combined with descriptive rDNS names and a dedicated abuse address. Second, we maintain a blacklist of hosts and networks which will not be scanned in any of our measurements. Third, we reply to every e-mail received on the abuse address. Throughout this experiment, we were not contacted by affected users.

As we only send `ICMP echo request` probes at a low rate, and use a 6-hour back-off per IP, we consider these measurements of low interference, as underlined by the fact that no users contacted us.

To conclude, we argue that no individual was harmed as a result of our active measurements. To avoid revealing private information, we take the following measures when releasing data: scans towards a public hitlist can be released in full, scans based on passive observations will be anonymized as suggested by Xu et al. [28] before release.

**Passive Measurements** offer potential for intrusion into user privacy. We protect privacy by only observing incoming traffic, by not storing any payload data, and by only capturing the external IP address of a communication. This prevents reconstruction of communication partners and content.

## 10. CONCLUSION AND FUTURE WORK

We deliver promising ground work on carrier-grade anomaly detection by using TTL values: We capture and exhaustively analyze the first of its kind carrier-grade TTL data set, finding a majority of IP addresses to be TTL-stable. For unstable IP addresses, we develop and quantify tractable subgroups and further analyze their TTL and Hop Count behavior. We correlate our data set with BGP data and find a clear, but unpredictive correlation. Comparison with pingback scans proves that these typically yield valuable insight into determining an anchor Hop Count for multi-TTL IP addresses. Hitlist scans into the Internet confirm that our passive data carries largely unbiased Hop Count data. We find variable portions of different subnet sizes to show uniform Hop Count behavior. These insights will be of great value in designing a state-of-the-art filter for carrier-grade TTL anomaly detection.

**Future Work:** We see potential for future work to (a) dissect TTL stability along additional criteria, e.g. server/client port, (b) quantify the impact of routing changes into TTL behavior, (c) further investigate and classify the network characteristics leading to several observed TTL values per IP, and (d) design an online TTL filter.

## 11. ACKNOWLEDGMENTS

We thank Christian Sturm, Arno Hilke and Till Wickenheiser for their contributions to the tools used. We thank Felix von Eye and the Leibniz Supercomputing Centre of the Bavarian Academy of Sciences and Humanities for their support, and Sebastian Gallenmüller for his comments.